\documentclass[english,a4paper,aps,prb,twocolumn]{revtex4}
\pdfoutput=1
\usepackage[utf8]{inputenc}
\usepackage{amsmath}
\usepackage{color}
\usepackage{graphicx}
\usepackage{float}

\begin{document}
\title{Imaging the localization of the quasi-bound states in graphene antidots}
\author{A. Mre\'nca, K. Kolasi\'nski, and B. Szafran}
\address{AGH University of Science and Technology, Faculty of Physics and Applied Computer Science,\\
al. Mickiewicza 30, 30-059 Krak\'ow, Poland}
\begin{abstract}
We consider charge transport across a finite graphene flake with a circular antidot
defined in its center. The flake is connected to thin metallic armchair nanoribbons and the study covers the energy range within the neighborhood of the neutrality point.
We solve the scattering problem using the tight-binding Hamiltonian and find that conductance of the system is non-zero only
near narrow resonances which are quasi-bound in either the entire cavity or the antidot
itself.
We demonstrate that
the scanning probe technique can be used for the purpose
of determination whether the state is localized within the antidot or in the entire cavity.
We indicate that the potential of the perturbation shifts the position of the resonances
and that the shifts are related to the localization of the scattering probability density.
The resonance lifetime can be both decreased or increased as the perturbation introduced by the probe interferes with the current vortices
inside the antidot.
\end{abstract}
\maketitle
\section{Introduction}
Graphene band structure -- gapless and linear near the neutrality point --
excludes electrostatic confinement of charge carriers \cite{kt,kk} in bound states
due to the Klein tunneling with electron-hole conversion at potential steps.
Nevertheless, an external local potential of a quantum dot or antidot introduced to the graphene plane allows
for formation of quasi-bound states\cite{silvestrov07,matulis08,apalkov08,badarson09,pal11,feshke13,feshke13epl}
which in the transport experiments produce Fano resonances of conductance by interference with the incident currents.
The Klein tunneling is strongly anisotropic \cite{kt} with the transfer probability reaching 100\%
only for some electron incidence angles to the potential step, the normal one in particular.
Formation of long living resonances in circular antidots with currents flowing tangential
to the edge was reported.\cite{badarson09}
The transverse motion of the carriers inside graphene quantum wells assists in formation of bound states for a range of wave vectors and energies,\cite{pereira06}
allowing for fabrication of Fabry-P\'erot interferometers with multiple internal reflections.\cite{fp}

In this work we study a finite flake of graphene containing  an antidot defined in its inside with  metallic armchair nanoribbons\cite{nanoribbons}
feeding the current to the system.
We evaluate the electron transfer probability solving the electron scattering problem \cite{tworzydlo}
for the electron incident from the input lead of a metallic armchair nanoribbon using the tight-binding Hamiltonian.
The conductance exhibits narrow peaks due to the resonant quasi-bound states.
The states are localized within the entire flake -- playing a role of a resonant cavity -- or inside the antidot.
The conductance dependence on the Fermi energy by itself does not allow for determination
whenever the state is localized within the antidot or the cavity.
We discuss the possibility of determination of the electron localization in the antidot by using the scanning gate microscopy.

The scanning gate microscopy \cite{sgmr} is a technique that introduces a local perturbation of the potential landscape
by the the atomic force microscope that is capacitively coupled to the sample. The scanning gate microscopy
of graphene-based systems was used  for universal conductance fluctuations,\cite{b1} probing
weak localization effects,\cite{b2} and the charge inhomogeneity.\cite{b3}
The technique allows for spatially resolved detection of the localized states in constrictions,\cite{c1} charge islands
due to the substrate-induced potential,\cite{c2} as well as for quantum dots formed by local widening of a nanoribbon.\cite{c3}

We look at the response of the system to the perturbation
by a short range potential simulating the probe scanning the surface of the system.
Since the conductance resonances are very narrow, the tip generally reduces the conductance down to zero.
We show however, that the maps of the energy shifts that the resonances undergo as functions of the tip position indicate very clearly the localization of the quasi-bound
states inside the antidot or in its surroundings within the flake. Moreover, for strong antidot potentials, also the
overall form of the resonant probability distribution within the antidot can be extracted from the map of the shifts.
We find that the perturbing potential deflects the charge currents and
largely modifies the resonance lifetime.

\section{Theory}

\begin{figure}[htbp]
 \includegraphics[width=0.35\textwidth]{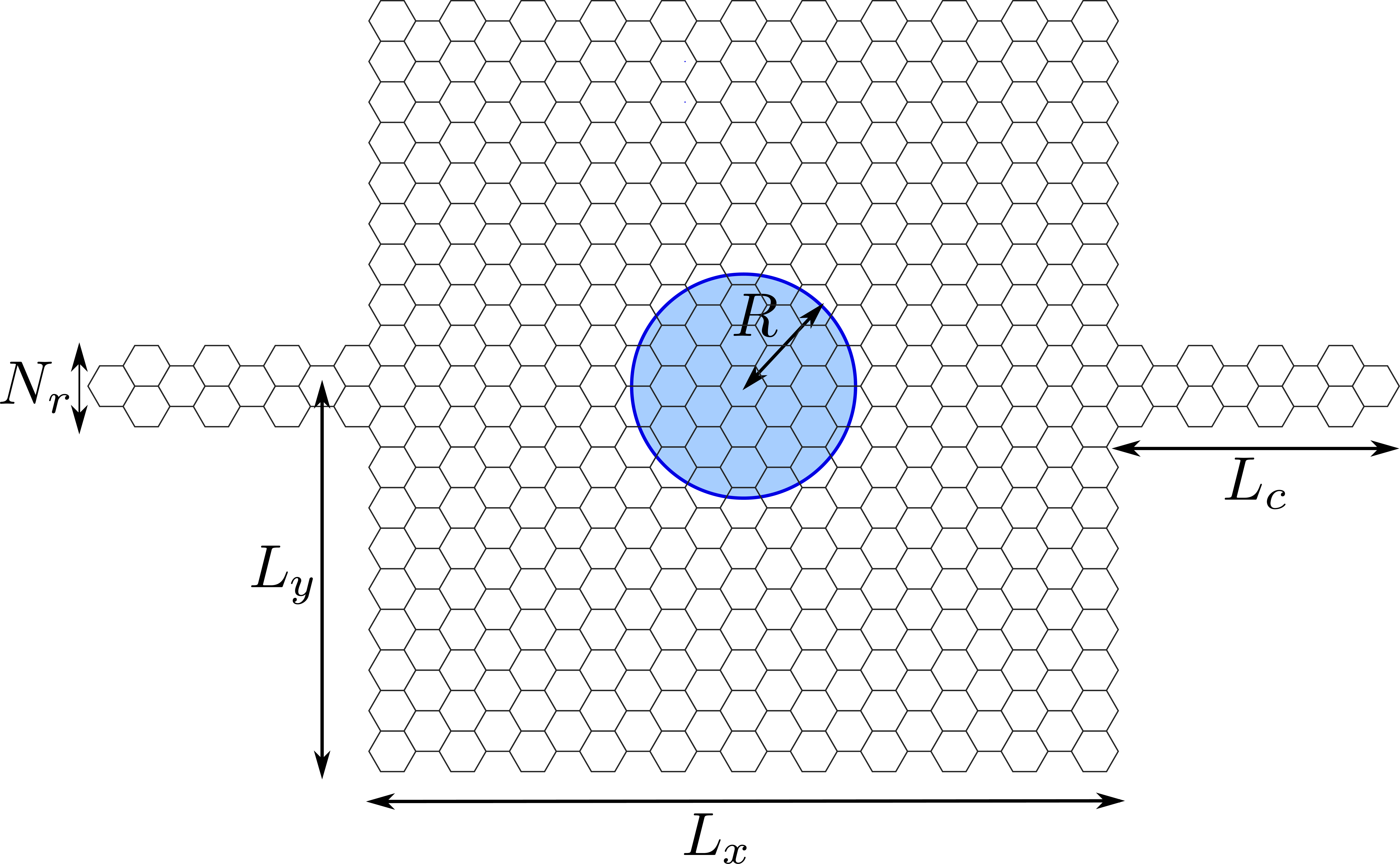}
\caption{Schematic drawing of the studied graphene system. Two armchair ribbons are connected to a flake
with centrally defined antidot of radius $R$. The ribbons are assumed of an infinite length. The computational box
covers the leads segment of length $L_c=153.36$ \AA . The flake has dimensions $2L_y = 398.52$ \AA, $L_x = 680.32$ \AA , $N_r=5$ atoms, unless
stated otherwise. The radius of the antidot is $R=40a$, where $a=2.46$ \AA$\;$   is the graphene lattice constant. }\label{lat}
\end{figure}

We consider the flake of a size of about $400$ \AA  $\times 680$ \AA $\;$ that is schematically depicted in Fig. \ref{lat} with zigzag vertical and armchair horizontal edges.
Narrow metallic armchair ribbons\cite{nanoribbons}  are connected to the flake.
 Inside the flake an antidot of radius $R=98.4$ \AA $\;$ is defined by e.g. external gate. The antidot potential is taken in form $V_s({\bf r})=V_0\theta(R-|{\bf r}-{\bf r_0}|)$, where ${\bf r_0}$ is the center of the antidot and $\theta$ is the Heavyside function.
Throughout the paper we use the tight-binding Hamiltonian
\begin{eqnarray}
H=\sum_{\{i,j\}}t_{ij} (c_i^\dagger c_j+c_j^+c_i)+\sum_i V({\bf r}_i) c_i^\dagger c_i, \label{dh}
\end{eqnarray}
where the first summation runs over the nearest neighbors with $t=-2.7$ eV.

\begin{figure}[htbp]
 \includegraphics[width=0.35\textwidth]{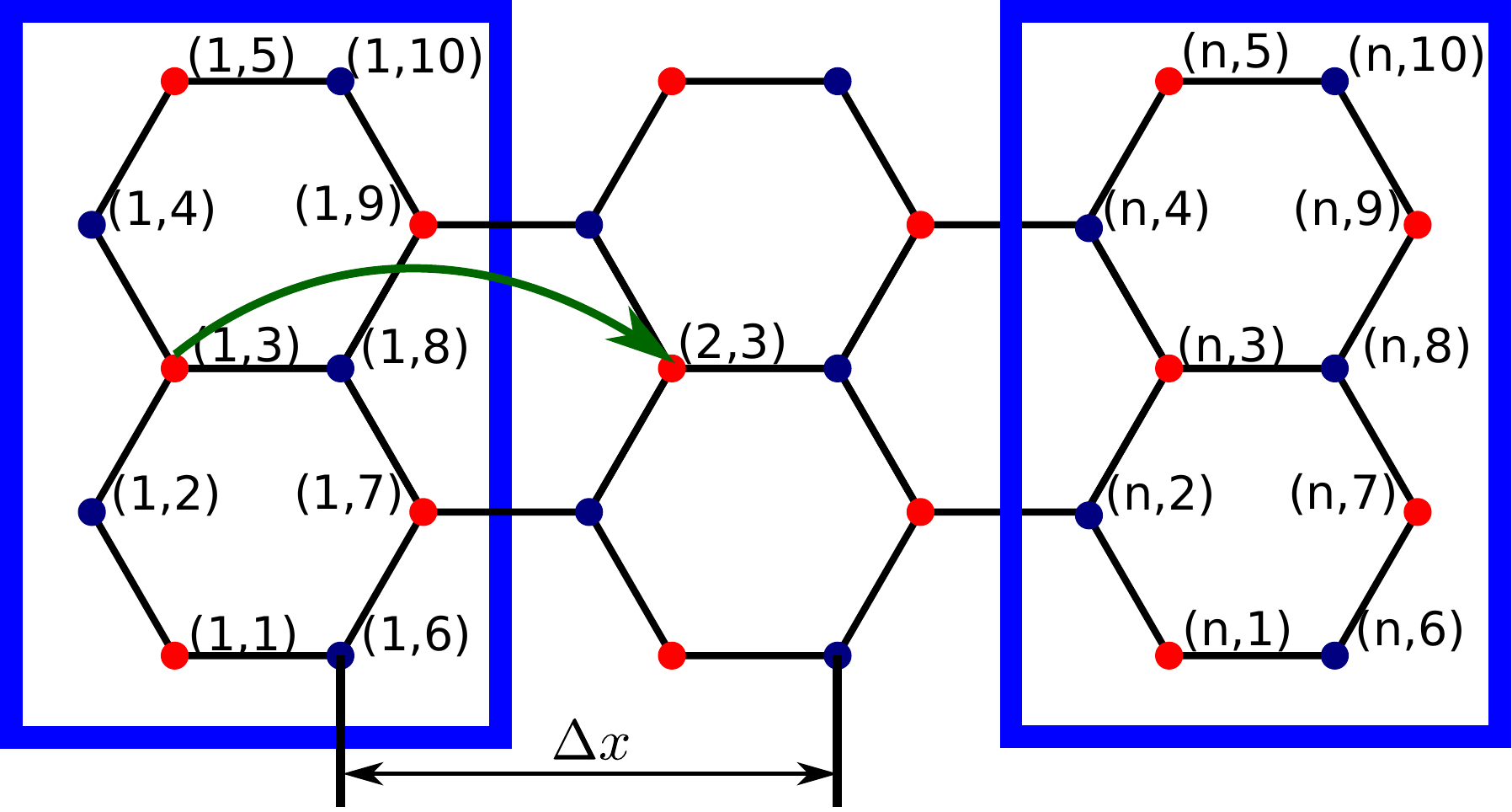}
\caption{A fragment of the nanoribbon with the elementary cell (blue rectangle) and the boundary conditions for the Bloch waves [see Eq. (\ref{bloch})].
$\Delta x$ is the length of the elementary cell. The basis atom $j$ in the elementary cell $i$ is denoted by $(i,j)$. For atoms with same $j$ the values of function $\phi^k(x,y)$ [Eq. (\ref{bloch})] are the same for all $i$. }\label{rdpng}
\end{figure}

We assume that the nanoribbon channels (see Fig. 1) have an infinite length. The electron eigenstates can be described by functions of Bloch periodicity
\begin{equation}
\Psi(x,y)=\exp(ikx) \phi^k(x,y), \label{bloch}
\end{equation}
where $k$ is the wave vector and $\phi^k$ is a periodic function $\phi^k(x+\Delta x,y)=\phi^k(x,y)$, where $\Delta x$ is the length
of the elementary cell -- see Fig. (\ref{rdpng}). The dispersion relation $E(k)$ and  $\phi^k$ functions are determined
by  diagonalization of the Hamiltonian (\ref{dh}) within an elementary cell.
Let us denote the atom $j$ in the elementary cell $m$ by $(m,j)$.
For the neighbors $(m,j)$ of the atom $(l,i)$ which are outside the elementary cell $l$ (blue
rectangle in Fig. \ref{rdpng})  we use the Bloch periodicity, $\Psi_{(m,j)}=\exp(i k (x_{(m,j)}-x_{(l,j)}))\Psi_{(l,j)}$, where
$x_{(m,j)}-x_{(l,j)}=\pm \Delta x$.

\begin{figure}[htbp]
 \includegraphics[width=0.34\textwidth]{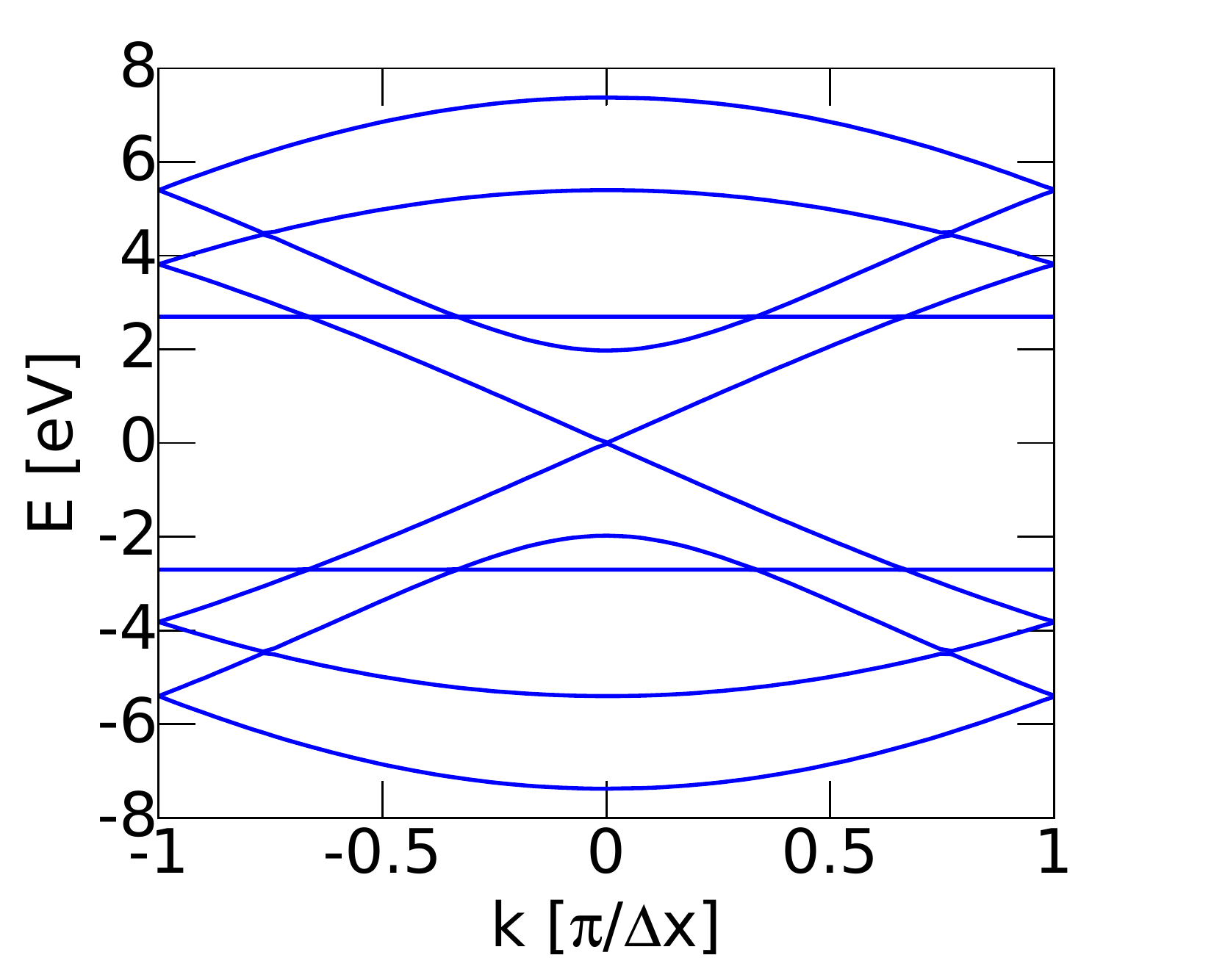}
\caption{The dispersion relation for the nanoribbon of Fig. \ref{rdpng}.}\label{dr}
\end{figure}

 In the present work we describe the resonances localized at the antidot and their imaging by the scanning gate microscopy.
We consider quantum transport for Fermi energies near the neutrality point. For the applied choice of the armchair (and not zigzag) nanoribbon there are no localized states at the
edges near the neutrality point.
The dispersion relation for 5 atoms across the ribbon (see Fig. 2) is displayed in Fig. \ref{dr},
with no energy gap that appears for semiconducting armchair ribbons (for multiple of 3 atoms across the channel).
 The metallic ribbon that we consider here carries the current in a single subband only, in a wide range of Fermi energies, $|E_F|<1.9\; \mathrm{eV}$.
 For wider ribbons this energy range is thinner -- for 17 instead of 5 atoms across the channel the single band transport occurs for $|E_F|<0.77\; \mathrm{eV}$.

In the input channel far away from the flake we have an incident electron wave function with $k>0$ superposed with the backscattered one with $k<0$,
\begin{equation}
\Psi(x,y)=c_{+}\exp(ikx) \phi^{k}(x,y)+c_{-}\exp(-ikx) \phi^{-k}(x,y). \label{inside}
\end{equation}
In the output ribbon we have only the outgoing wave function
 \begin{equation}
\Psi(x,y)=c_{out} \exp(ikx) \phi^{k}(x,y). \label{outside}
\end{equation}
For evaluation of conductance we solve the scattering problem for Hamiltonian (\ref{dh}) for a chosen Fermi energy $E$
and the boundary conditions given by Eqs. (\ref{inside},\ref{outside}). The scattering amplitudes $c_k^\pm$ and $c_{out}$ are found using an iterative scheme
similar to the one described in Ref. \onlinecite{szafran} for description of the Fermi level transport in 2DEG in GaAs.
After the convergence is reached the electron transfer probability is calculated as $T=|c_{out}/c_{k^+}|^2$.
For the single-subband transport the conductance is given by the Landauer approach\cite{tworzydlo} as $G=\frac{2e^2}{h}T$.

In the discussion we refer to currents flowing within the graphene flake. The currents in the tight binding approach
flow along the $\pi$ interatomic bonds, and the formula for the current flowing from atom $l$ to atom $j$ derived \cite{waka} from the Schr\"odinger equation
is \begin{equation}
 J_{lj} =\frac{i}{\hbar} \left[ t_{lj} \Psi^*_l \Psi_j - t_{jl} \Psi_l \Psi^*_j \right].
\end{equation}
For discussion of the current distribution we show values that are averaged over square cells of area 8\AA $\times$ 8\AA.

The charged tip of the atomic force microscope interacts with the electron gas which induces re-distribution of the electron density and
as a consequence screens the Coulomb tip potential. The form of the effective tip potential was derived from the Schr\"odinger-Poisson modeling of the SGM
for a two-dimensional electron gas in Ref. \onlinecite{szafran}. It was found \cite{szafran} that the screened potential is short-range and can be quite
accurately modeled by a Lorentz function, which we apply in this work
$$ V=\frac{U_{tip} d^2}{ (x-x_0)^2+(y-y_0)^2+d^2 }. $$
The width of the tip $d$ is nearly equal to the distance between the tip and the two-dimensional electron gas and is insensitive \cite{szafran} to both potential (charge) at
the tip and the density of the two-dimensional electron gas. The tip voltage can be freely varied and the carrier density in graphene is strongly dependent on
the sample fabrication details. In this work we consider $U_{tip}$ in a wide range between 10 and 100 meV and find that the maps of the resonance energy shifts preserve their pattern as $U_{tip}$ is varied.
Only the amplitude of the maps change with $U_{tip}$ (see below). For the width of the tip we take $d=R/8$, which for the present value of $R$ gives the width of the tip potential equal to 12.3 \AA$\;$which is the shortest surface-tip distance for AFM operation in the non-contact mode.
Naturally, for larger $d$ resolution of the energy maps is reduced.

\begin{figure}[htbp]
\includegraphics[width=0.5\textwidth]{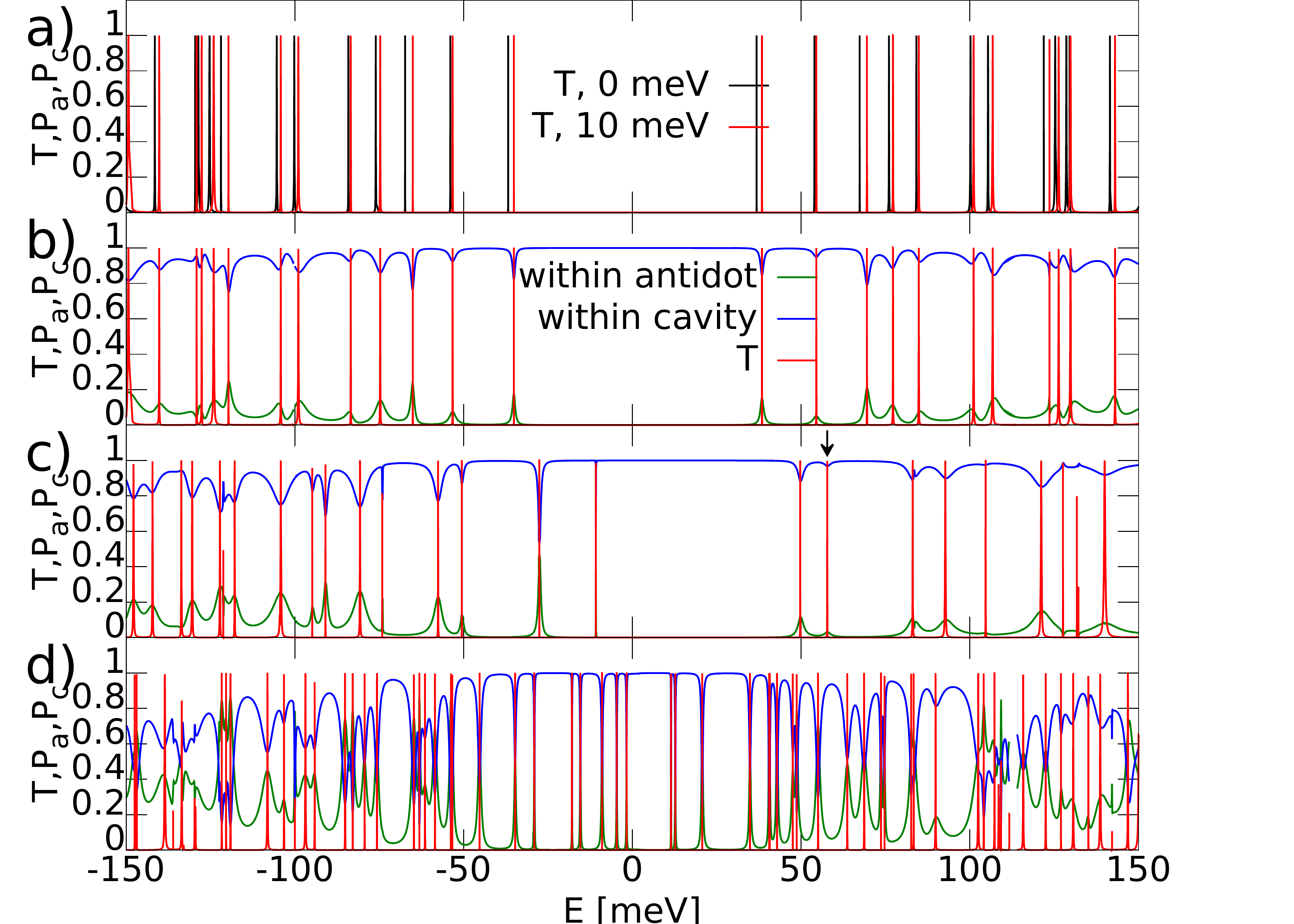}
\includegraphics[width=0.5\textwidth]{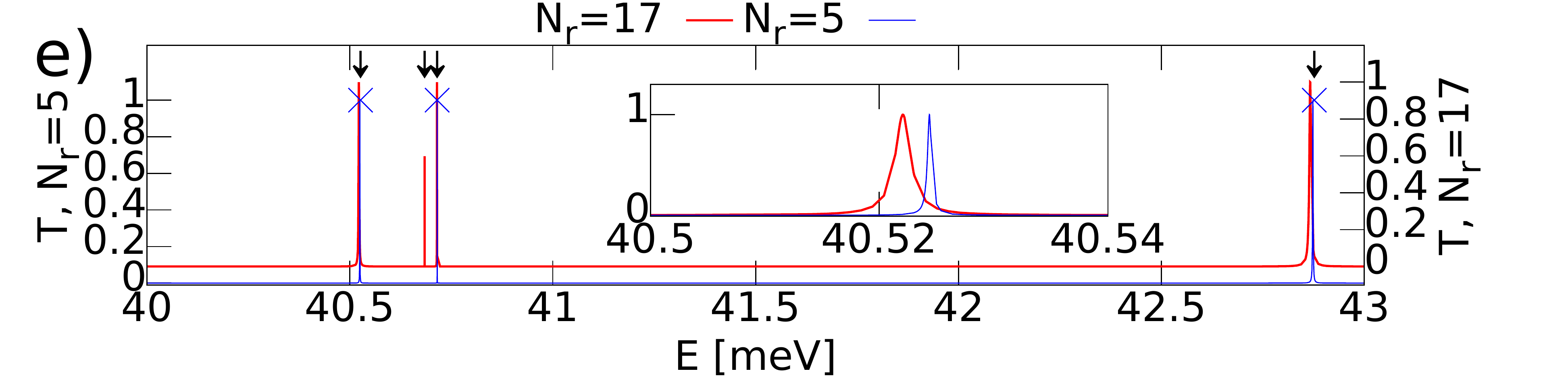}
\includegraphics[width=0.5\textwidth]{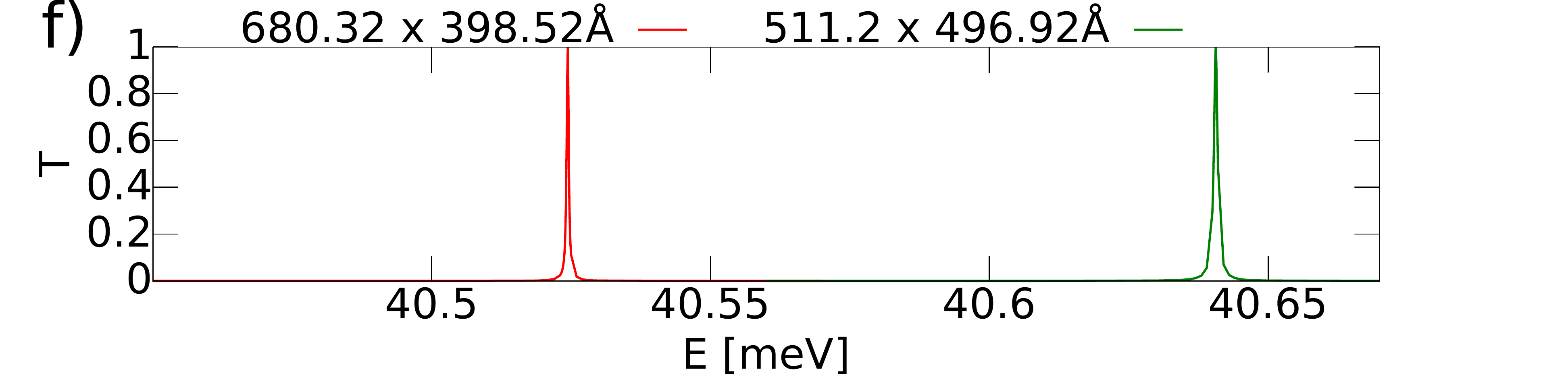}
\caption{a) Transfer probability as a function of the energy for the empty cavity (black lines) and the cavity
with the shallow antidot $V_0=10$ meV (red lines).  b) The shallow cavity with the antidot with $V_0=10$ meV. The red lines show the transfer probability. The green (blue) curves correspond to the part of the probability density that is localized inside the antidot $P_a$ (within the cavity but outside the antidot $P_c$). For the plot, the probability density is normalized to 1 within the cavity.
Results for c) and d) correspond to $V_0=100$ meV and $V_0=1$ eV, respectively.
 e) Transfer probability for leads with $N_r=5$ and 17 atoms. The arrows indicate the energies calculated for cavity uncoupled to the leads.
f) Comparison of the resonance in two cavities with different sizes.
}
\label{wn}
\end{figure}

\begin{figure}[htbp]
\includegraphics[width=0.35\textwidth]{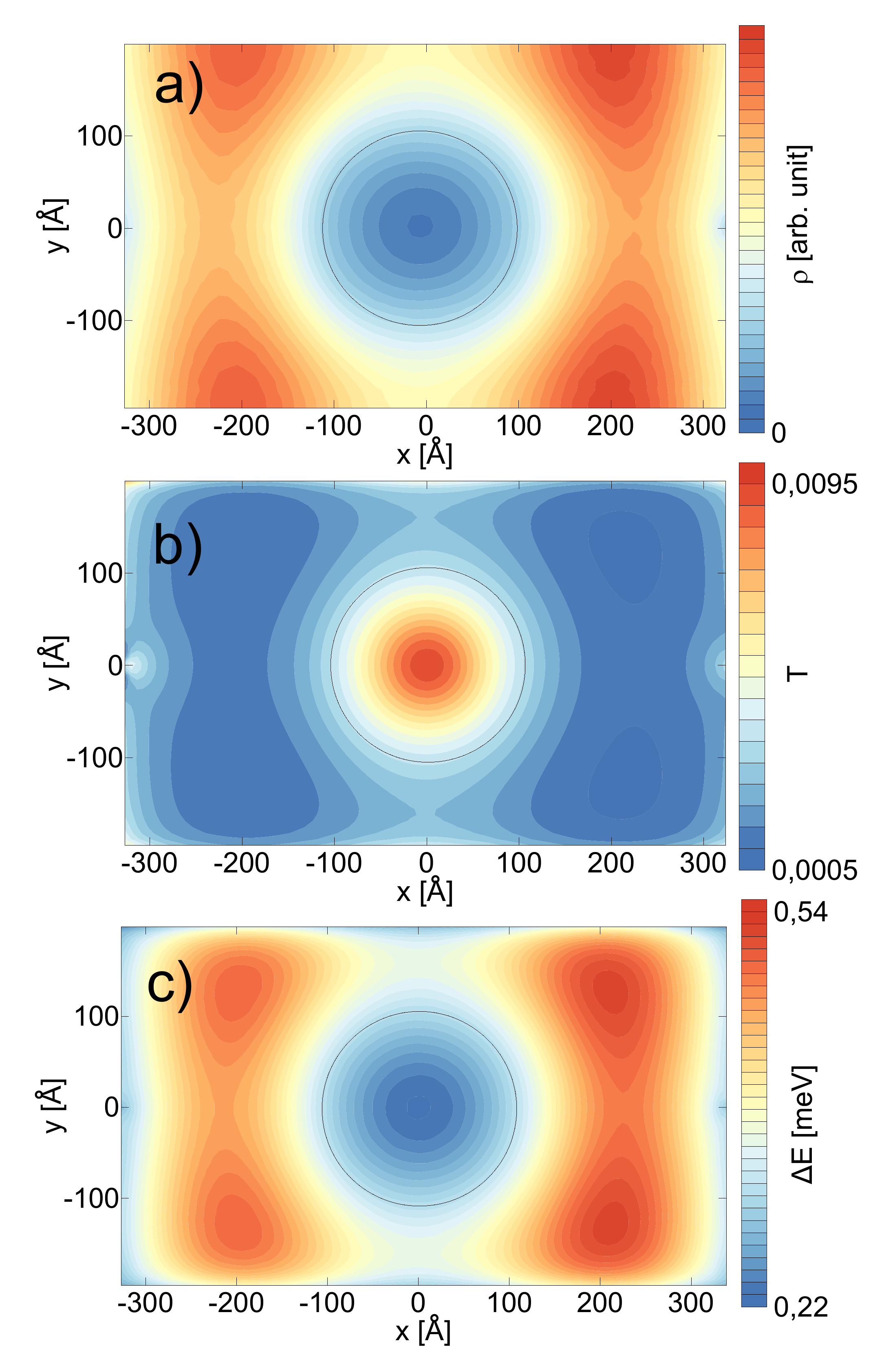}
  \caption{a) Probability density for the peak of $T(E)$ at $E=57$ meV for the antidot with potential $V_0=100$ meV -- see the peak marked by
the arrow at Fig. \ref{wn}(c). b) The electron transfer probability for the tip with $U_{tip}=100$ meV scanning the surface of the sample
as a function of the tip position. c) Shift of the resonance position as a function of the tip position.}
\label{fwn}
\end{figure}

\begin{figure}[h!]
 \includegraphics[width=0.5\textwidth]{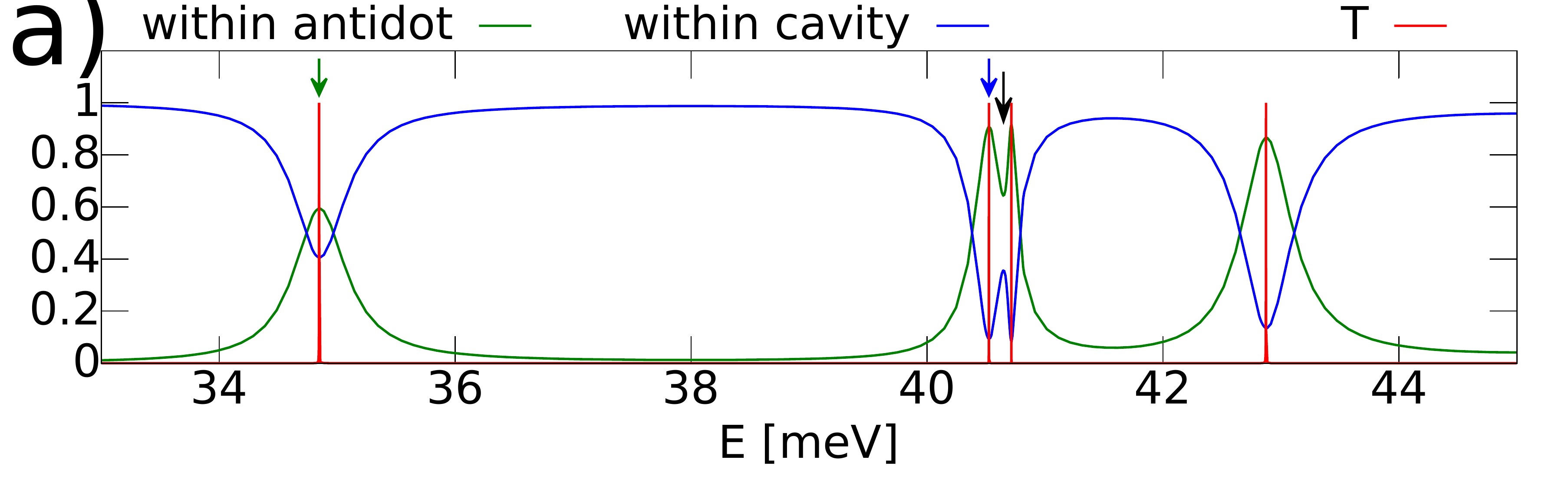} \\
 \includegraphics[width=0.5\textwidth]{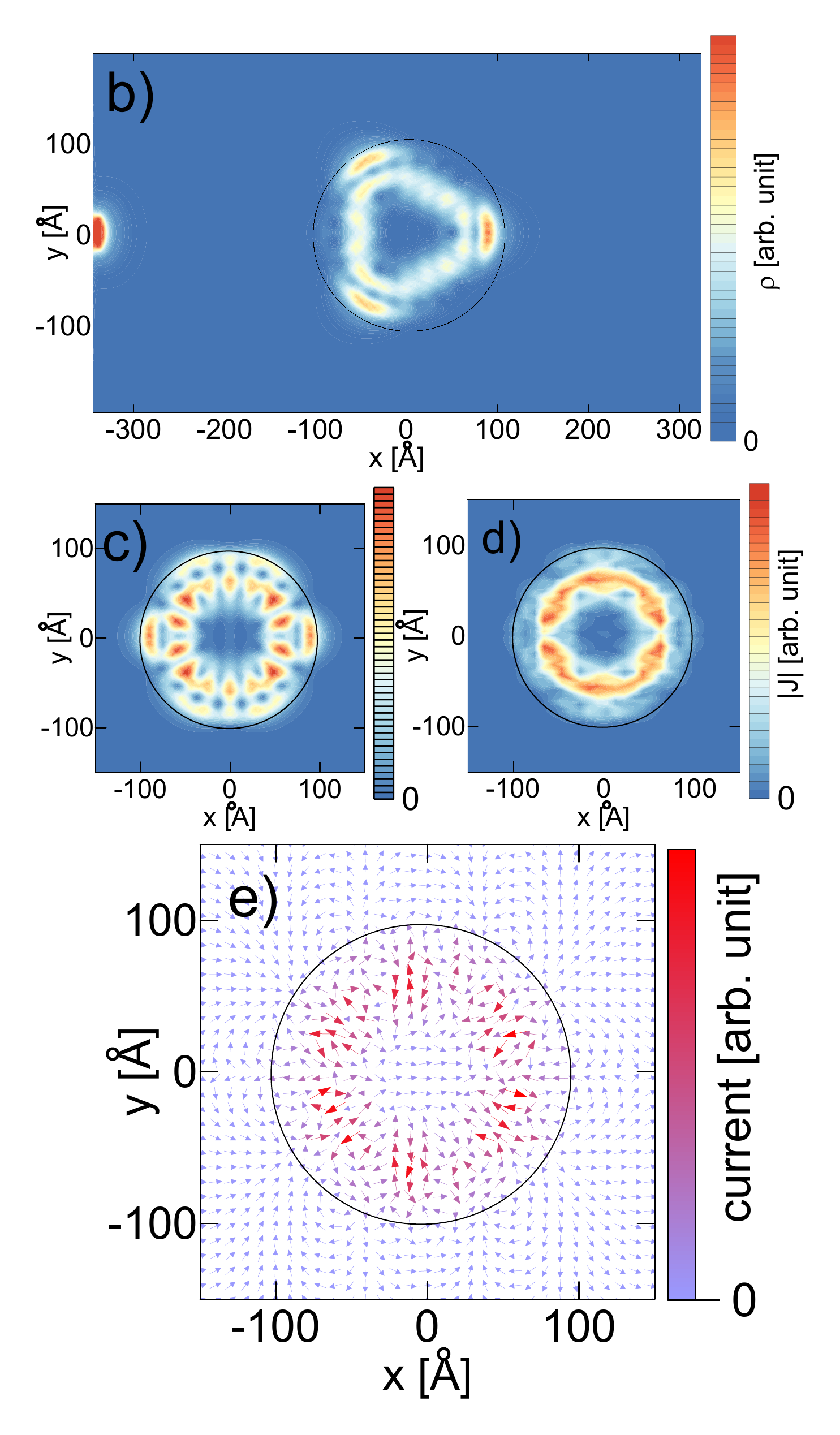} \\
  \caption{a) Zoom of Fig. \ref{wn}(d) , b) The probability density for $E=40.65$ meV (for $T=0$) marked in (a) by the black arrow.
  The probability density (c) current density (d) and current map (e) at the resonance $E=40.5$ meV  -- the blue arrow in (a).
  The colors in the probability density maps plotted in (b) and (c) use different scales. Same applies for the rest of the probability density plots in
  this work.}
\label{res}
\end{figure}

\begin{figure}[htbp]
 \includegraphics[width=0.5\textwidth]{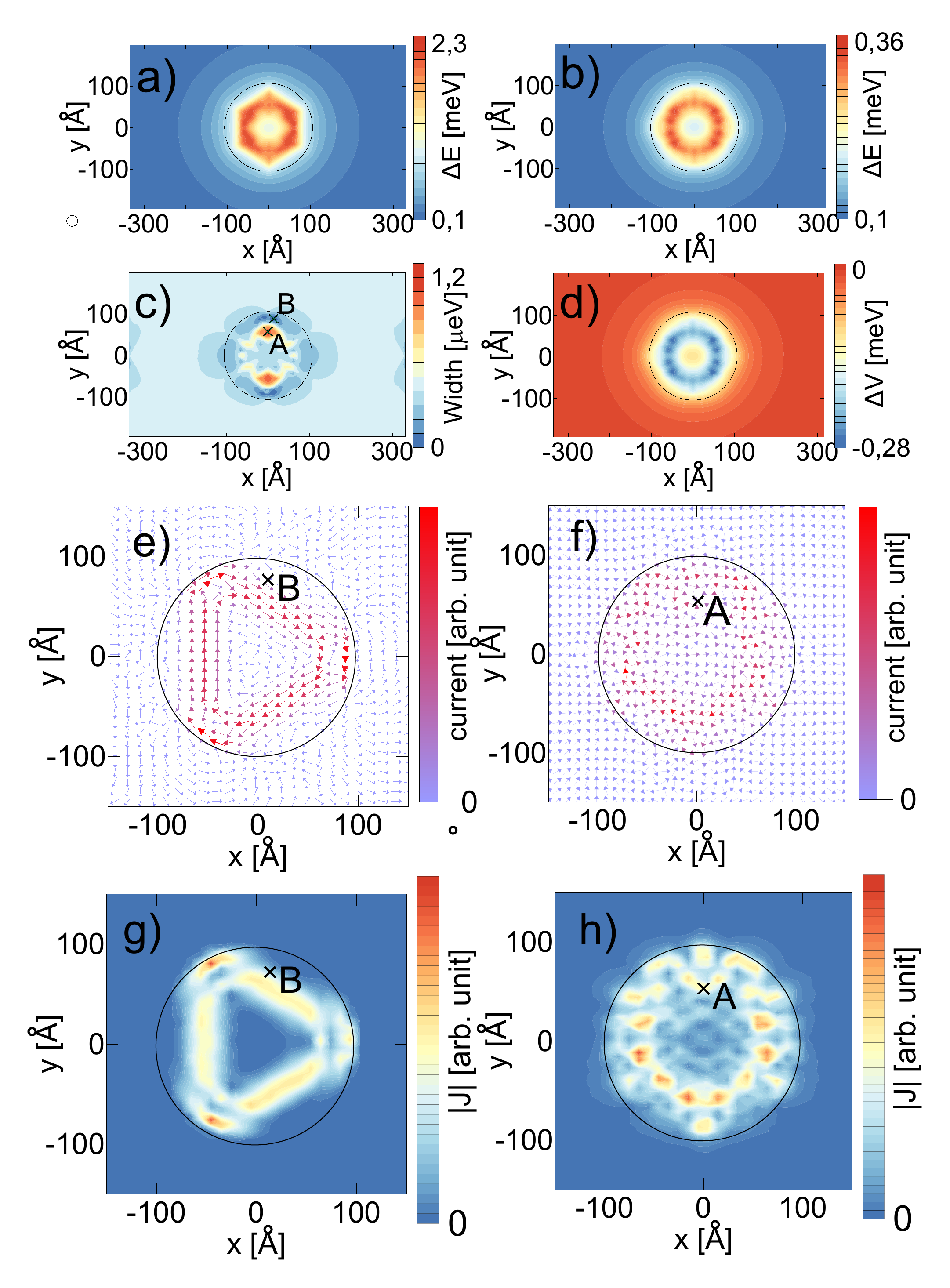}
\caption{Results for the resonance at $E=40.5$ meV -- see the  blue arrow in Fig. \ref{res}.  Shifts of the peak  position
induced by the tip above point $x,y$ for $U_{tip}=100$ meV (a) and $U_{tip}=10$ meV (b). The width of the shifted peak
for $U_{tip}=100$ meV (c) as a function of the tip position. Variation of the potential depth necessary to keep the resonance
at $E=40.5$ meV as a function of the tip position for $U_{tip}=10$ meV (d). (e) and (f) show the current flow map,
 and (g) and (h) - current density map for the tip above points B and A marked
in (c).
}\label{zbiorczy}
\end{figure}

\begin{figure}[htbp]
 \includegraphics[width=0.4\textwidth]{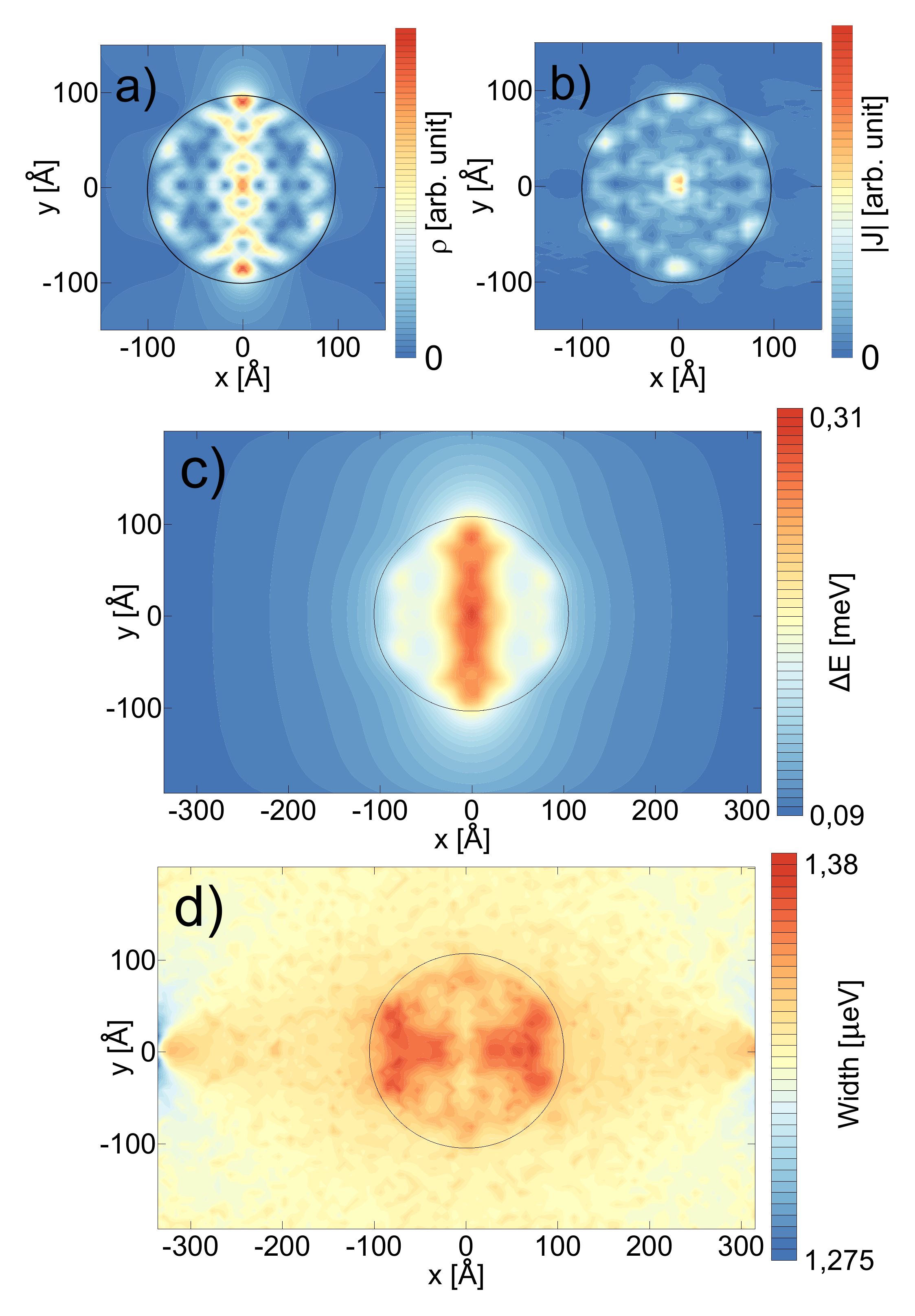}
\caption{Results for the resonance at $E=34.8$ meV -- see the green arrow in Fig. \ref{res}.  The scattering probability density (a),
the current density (b),
the shifts of the peak position (c) and the width of the resonances (d) as functions of the tip position for $U_{tip}=10$ meV.
}\label{zbiorczy2}
\end{figure}

\begin{figure}[htbp]
\begin{tabular}{ll}
\includegraphics[width=0.5\textwidth]{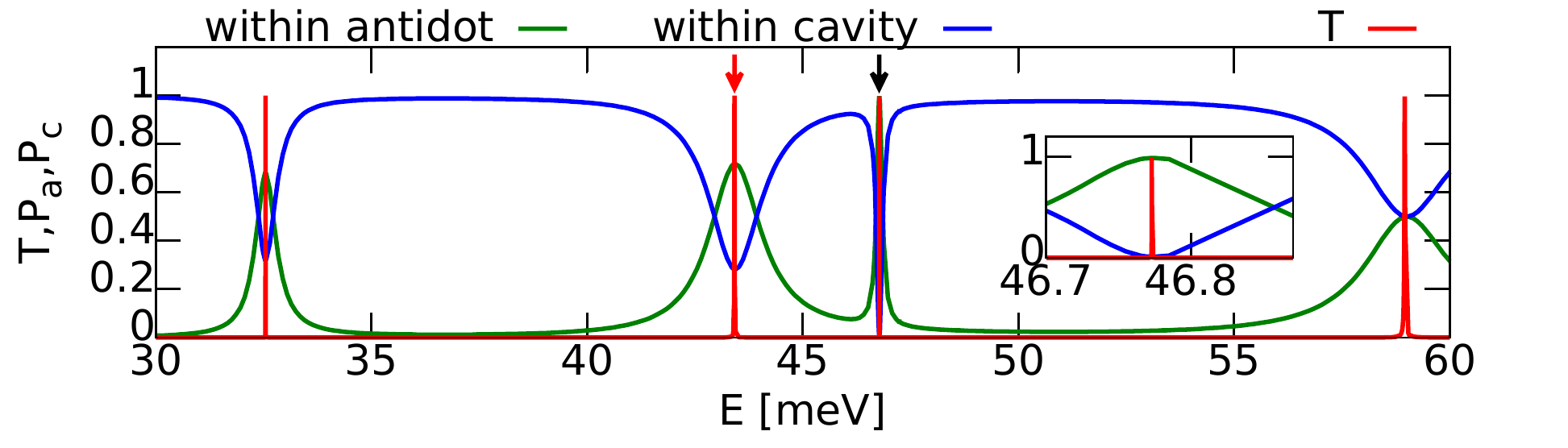}
  \end{tabular}
\caption{ Transfer probability as a function of the energy for the smooth cavity with $V_0$ = 1
eV (red lines). The green (blue) curves correspond to the part of the probability density that is localized inside the antidot (within the cavity but outside
the antidot). The probability density is normalized to 1 within the cavity. The arrows indicate the resonances that are discussed in the text.
The inset shows the zoom around the  peak position  that is marked with the black arrow.
}\label{Tgl}
\end{figure}

\begin{figure}[htbp]
 \includegraphics[width=0.5\textwidth]{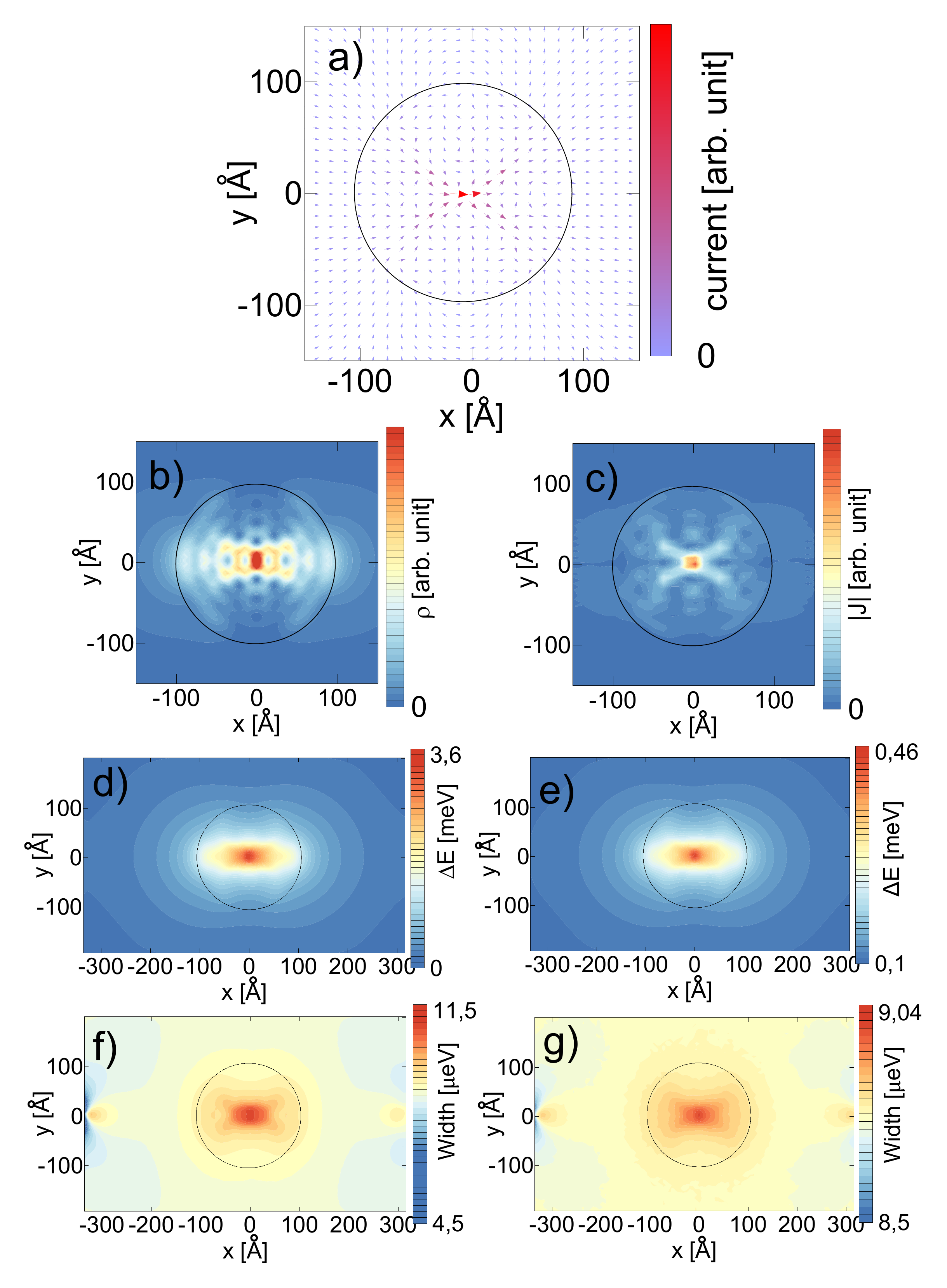}
\caption{Results for the resonance at $E=43.4$ meV -- see the red arrow in Fig. \ref{Tgl}.
(a) The current flow map,
 (b) The probability density, (c) current density map at resonance. (d) and (e) show the shifts of the peak position with $U_{tip}=100$ meV and $U_{tip}=10$ meV, respectively.
(f) and (g) show the width of the shifted peak with $U_{tip}=100$ meV and $U_{tip}=10$ meV, respectively.
}\label{zbiorczy2s}
\end{figure}

\begin{figure}[htbp]
 \includegraphics[width=0.5\textwidth]{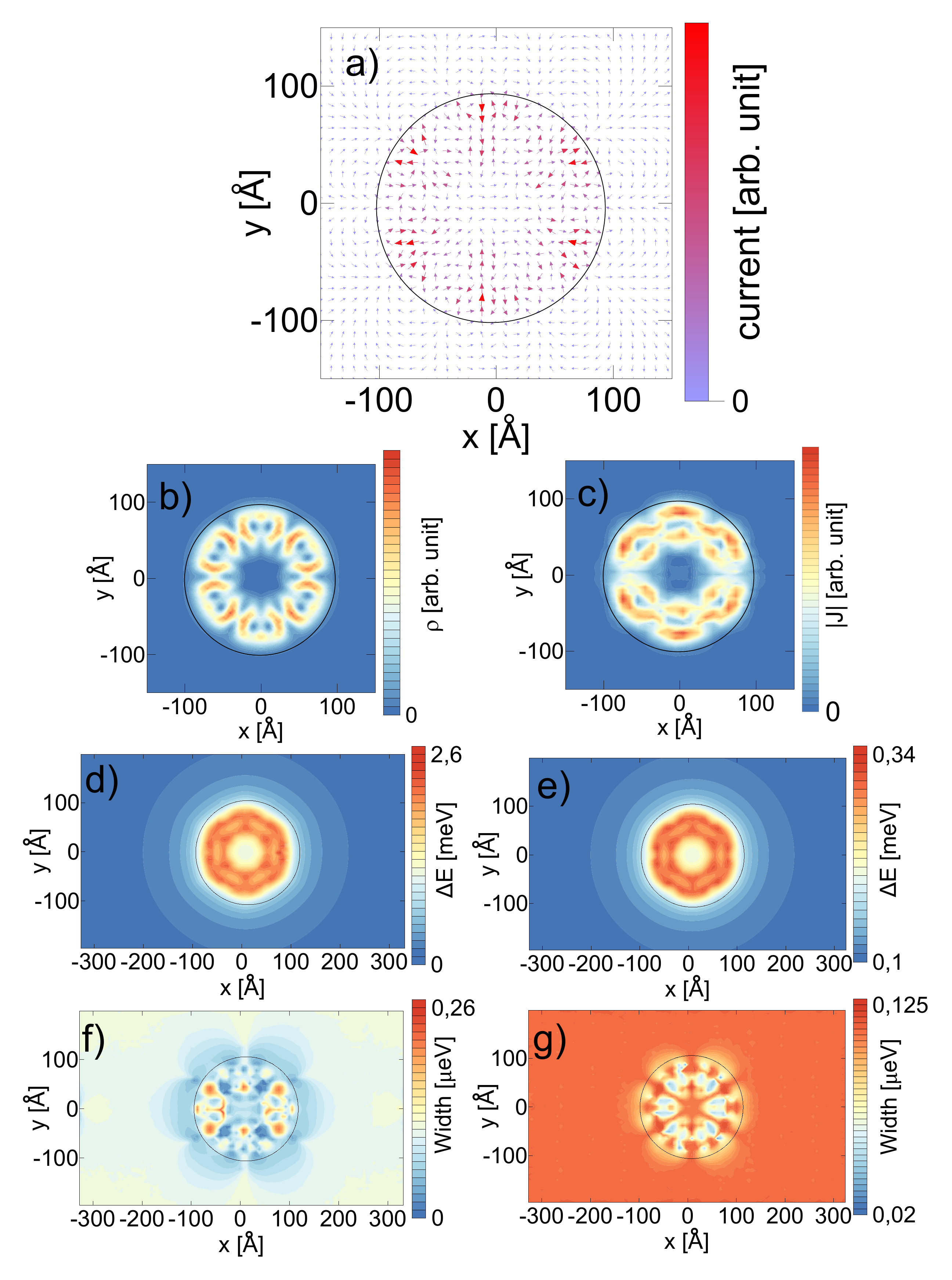}
\caption{Results for the resonance at $E=46.8$ meV -- see the black arrow in Fig. \ref{Tgl}.
(a) The current flow map,
 (b) The probability density, (c) current density map at resonance. (d) and (e) show the shifts of the peak position with $U_{tip}=100$ meV and $U_{tip}=10$ meV, respectively.
(f) and (g) show the width of the shifted peak with $U_{tip}=100$ meV and $U_{tip}=10$ meV, respectively.
}\label{zbiorczy3s}
\end{figure}

\section{Results}

Figure \ref{wn}(a) shows the electron transfer probability for the empty cavity (black lines).
Generally the cavity is opaque for the incident electrons unless the Fermi energy
coincides with the localized resonances (quasi-bound energy levels) of the cavity.
When the antidot of $V_0=10$ meV is introduced to the system the symmetry of $T(E)$ with respect to the neutrality point is lifted, and the transfer probability peaks are shifted to higher energies by
a few meV [see Fig. \ref{wn}(a)].
In Figure \ref{wn}(b) we plotted the part of the scattering probability density localized inside the antidot (green lines) and the rest
of the cavity (blue lines) for $V_0=10$ meV.
At the resonances the probability density reaches the antidot, hence the maxima of the green lines in Fig. \ref{wn}(c),
but still most of the electron density is localized outside the antidot.

Generally the resonances for $V_0=10$ meV and for $V_0=100$ meV [Fig. \ref{wn}(b) and (c)] correspond to states quasibound inside the entire cavity and
not on the antidot.
Figure \ref{fwn}(a) shows a typical probability density for a resonance at $E=54$ meV and $V_0=100$ meV [see the arrow in Fig. \ref{wn}(c)].
We studied the reaction of this resonance to the tip potential scanning the surface above the sample. The transfer probability dependence
on the tip position is displayed in Fig. \ref{fwn}(b). Since the peak of $T(E)$ is very sharp,\cite{pal11} the tip presence
strongly reduces the conductance. This abrupt reduction is the smallest for the tip above the antidot, which
is consistent with the minimum of the probability density distribution found at the antidot [Fig. \ref{fwn}(a)].
We looked for the shifts of the incident electron energy $\Delta E$ that are needed to restore the resonance condition $T(E+\Delta E)=1$.
The map of calculated energy shifts of the resonances is displayed in Fig. \ref{fwn}(c) as a function of the tip position.
In accordance with the results of Fig. \ref{fwn}(a,b) we notice that
the reaction of the resonances to the tip position is the weakest for the tip above the antidot.\cite{uwaga}
Concluding for a shallow antidot the resonances are typically localized outside the antidot.
Both maps of $T(x,y)$ and $\Delta E(x,y)$ clearly indicate this type of localization.

In Fig. 4(e) we plotted the fragment of $T(E)$ plot of Fig. 4(d) for $N_r=5$ atoms across the input and output lead
(as elsewhere in this paper) and the $T(E)$ values for a wider channel with $N_r=17$ atoms across. The width of the peaks
is increased for the wider channels and only a small shift of the positions of the peaks is observed.
The discussed resonances are weakly coupled to the channels. For comparison by arrows we marked the energy
levels for the cavity with the antidot presence that is obtained for the leads removed. A very good coincidence
of the $T(E)$ peaks with the energies of the bound states is observed. Note, that the second energy level from the left
is only resolved in $T(E)$ for $N_r=17$.

For steeper potential, we observe a stronger localization of the resonances inside the antidot [see the green curve in Fig. \ref{wn}(d) for $V_0=1$ eV].
In Fig. \ref{res}(a) we enlarged a fragment of Fig. \ref{wn}(d). The $T(E)$ dependence remains very sharp also at this scale. Figure \ref{res}(b) shows
a typical probability density distribution off the resonance for $E=40.65$ meV --- see the black arrow in Fig. \ref{res}(a).
The electron density penetrates the inside of the flake -- see the spot of larger density
at the left hand side -- and the localized state inside the antidot is found.
The incident electron in this case is still backscattered with a 100\% probability.
The current flux calculated along any line perpendicular to the axis of the channel  is zero.

In Fig. \ref{res}(c) the probability density at the exact position of the resonance $E=40.5$ meV (see the blue arrow in Fig. \ref{res}(a)) is plotted.
Now, we have a constant nonzero flux of the current. The current is strongly amplified within the antidot [Fig. \ref{res}(d)] as compared
to the cavity area outside the antidot. Within the antidot the current forms loops at a distance from
the antidot edge [Fig. \ref{res}(e)]. The discussed resonance is nearly entirely localized within the antidot [see the green curve at the place marked by the blue arrow in Fig. \ref{res}(a)].
In Fig. \ref{zbiorczy}(a,b) we plotted the shifts of the resonances induced by the tip scanning the surface of the sample for this resonance,
for $U_{tip}=100$ meV (a) and $U_{tip}=10$ meV (b) with  resolved resonance localization in the antidot.
We find an overall similarity between the scattering probability distribution [Fig. 6(c)]
and the maps of the shifts [Fig. \ref{zbiorczy}(a,b).
We find that in general the maps of the energy shifts are very well correlated to both the probability density distribution in the absence of the tip. 
The unperturbed current field distribution $|J|$ is usually very similar to the probability density maps in the absence of the tip. Nevertheless, occasionally we find an exception to the latter rule. One of them  is displayed in Figure 8. We have a very distinct probability density map [Fig.8(a)] and the map of the current amplitude [Fig.8(b)]. The energy shifts as obtained with the tip reproduce the probability density map and not the current distribution. Summarizing, the scanning probe imaging that we consider here can be applied to read-out the scattering probability density distribution and not directly the current distribution.

In the absence of the tip the resonance discussed in Fig. \ref{zbiorczy} has the width of $0.4\mu$eV [Fig. \ref{res}(a)],
corresponding to the lifetime of the quasi-bound state of 1.65 ns.
We found that the tip not only shifts the resonances but also changes their width (lifetime) and in a very pronounced manner.
The map of the resonance width  is plotted in Fig. \ref{zbiorczy}(c) as a function of the tip position. The
extrema of the width are found for the tip above the points marked by $A$ and $B$ in Fig. \ref{zbiorczy}(c).
The current maps for the maximal width of the peak 1.2$\mu$ eV [see point A in Fig. \ref{zbiorczy}(c)]
is given in Fig. \ref{zbiorczy}(f).
The regular flow pattern that was found in the absence of the tip [Fig. \ref{res}(d)] is disturbed and
the resonance is destabilized with the reduction of the
lifetime from 1.65 ns to 0.55 ns.
The minimum of $B$ corresponds to width of $0.016 \mu$eV only, with the corresponding resonance lifetime as large as 41.6 ns.
This is quite a remarkable result: the lowered symmetry of the potential due to the tip makes the resonance more stable.
The current distribution for the tip position inducing the maximal lifetime of the resonance is given in Fig. \ref{zbiorczy}(e),
in which we find that the current distribution forms a triangular loop that is nearly ideally tangential
to the edges of the antidot, which results in the electron storage within the antidot resulting in the increase of the lifetime.
Ref. \onlinecite{badarson09} discussed an opposite phenomenon: destabilization of the resonances for stadium cavities with respect to the circular ones.

Figures \ref{zbiorczy}(a) and \ref{zbiorczy}(b) have a similar pattern indicating that the results are robust against the specific
value of the tip potential. Naturally, the energy shifts depend on the tip potential -- note that the scale of shifts is very
different in Figs. \ref{zbiorczy}(a) and \ref{zbiorczy}(b) calculated for $U_{tip}=100$ meV and  $U_{tip}=10$ meV, respectively.
The resonance width maps are quantitatively nearly identical for $U_{tip}=100$ meV [Fig. \ref{zbiorczy}(c)] and $U_{tip}=10$ meV (not shown).
In the experiments it should be easier to keep the incident (Fermi) energy constant and tune the potential of the antidot
to maintain the resonant conditions when the tip scans above the surface. The variation of the antidot potential
necessary to keep the resonant conditions is given in Fig. \ref{zbiorczy}(d) for $U_{tip}=10$ meV.
We can see an ideal correspondence of this result to the shift of the resonance for fixed $V$ given in Fig. \ref{zbiorczy}(b).

The lowest energy resonance of Fig. \ref{res}(a) [green arrow] corresponds to the probability density that is localized inside the antidot
only in about 50\%. The corresponding probability density obtained from the scattering problem is plotted in Fig. \ref{zbiorczy2}(a). The shifts of this resonance as a function of the tip position with $U_{tip}=10$ meV is given in Fig. \ref{zbiorczy2}(b). We notice, that the resonance
reacts to the tip position also when the tip is quite far away from the antidot [cf. Fig. \ref{zbiorczy}(b)] indicating
the leakage of the probability density to the cavity area. The extinction of the shift for the previous resonance -- fully localized inside the antidot -- was
distinctly faster [see Fig. \ref{zbiorczy}(b)].  Note, that the central vertical dash of the probability density of Fig. \ref{zbiorczy}(a) is
resolved by the map of the shifts [Fig. \ref{zbiorczy}(b)].  The overall variation of the resonance lifetime is less pronounced as compared to the result
for the resonance entirely localized within the cavity [cf. Fig. \ref{zbiorczy2}(c) and Fig. \ref{zbiorczy}(c)].

\subsection*{Antidot with a smooth potential}

The potential induced electrostatically within the graphene plane is bound to be smooth,
without a well defined boundary. Since the profile of the potential near  the boundary is naturally likely to affect
the localization of the quasi-bound states we performed calculations for a modified confinement potential using,
$V=V_0\exp(-(|{\bf r-r_0}|/R)^{10})$.

The results for the electron transfer probability for $V_0=1$ eV as  displayed in Fig. \ref{Tgl} are
qualitatively similar to the results for the step-like potential of Fig. \ref{wn}(d).
Below we consider two resonances of the fragment of Fig. \ref{Tgl}
that are marked by arrows.
 The resonances marked by the red and black arrow have the width (lifetime): 8.8 $\mu$eV (6 ns) and 0.1 $\mu$eV (75 ps), respectively,
with the part of the  probability  density that can be found in the antidot equal to 72\% and 99\%, respectively.

Figure \ref{zbiorczy2s} shows the probability density in panel (a),
and the current in panel (b) for the wider resonance at the energy of $E=43.4$ meV.
The resonance has a peculiar property of focusing the current in the center of the antidot.
A similar focusing (lensing) for the probability density was reported for large antidots in Ref. \cite{feshke13}
The current passes across the antidot boundary with a nearly normal incidence,
for which the Klein tunneling is most pronounced. The antidot boundary is nearly transparent for the electron
flow hence the low lifetime of the resonance states.
We found as a general rule that similar current paths that are focused in the center of the quantum dot potentials are found for  $T(E)$ peaks
have a width from 1 to 10 $\mu$eV and larger. In these resonances less than 70\% of the probability density
is localized inside the antidot. The other $T(E)$ peaks with width of the order of $0.1 \mu$ eV exhibit
a larger presence of the scattering density inside the antidot and current vortices circulating inside the quantum dot.
Concluding, the current lensing effect is found for antidot resonances that are more strongly coupled to the cavity.

The probe scanning the surface of the system finds the horizontal dash formed by the scattering
probability density [Fig. \ref{zbiorczy2s}(a)] in the maps of the shifts of the resonances [Fig. \ref{zbiorczy2s}(c,d)].
For both values of $U_{tip}$ considered we notice that the shifts are detectable for the tip
far away from the antidot, which is consistent with the relatively low value of the
probability density that is found in the antidot. The width of this resonance is only increased
by the perturbation [Fig. \ref{zbiorczy2s}(e,f)].  The width is most strongly enhanced for the tip localized at the center of the antidot,
exactly in the area where the current is focused.

The corresponding results for the resonance of the longer lifetime [black arrow in Fig. \ref{Tgl}] are displayed in Fig. \ref{zbiorczy3s}.
This resonances form current loops that circulate near the ends of the antidot where most
of the probability density is localized.
The map of the resonance energy shifts resolve the general form of the probability density inside
the dot. The energy shifts of the resonance disappear for a closer distance between the tip and the antidot as compared
to the results for the previous resonance, in consistence with the larger extent of electron localization within the antidot.
The resonance width is now increased or decreased depending on the specific position of the tip
with respect to the current loops, as seen previously for the resonance in Figures \ref{res} and \ref{zbiorczy}.

\section{Summary and Conclusions}

We considered the electron flow across a graphene flake with an antidot formed  in its center
by an external potential using the tight-binding Hamiltonian. The current flow across the system
through metallic nanoribbon leads has been calculated using the solution of the quantum scattering problem for the Fermi level electrons.
The resonances of the electron transfer probability are related to the states quasibound inside the flake
or inside the antidot for larger potential that defines it. We simulated mapping localization of the resonances
by the scanning probe measurement. We found that the maps of the energy shift of the resonances induced by the tip
allow for determination whether the state is quasibound in the antidot or within the entire flake. Moreover, for
large antidot potential the details of the scattering probability distribution within the antidot can be resolved.
We found that for resonances forming loops of current within the antidot the width of the resonances is a non-monotonic
function of the position of the tip, which interferes with the current vortices in a way that can stabilize
or destabilize the resonance depending on the tip position.


\begin{thebibliography}{00}
\bibitem{kk} M.I. Katsnelson, K.S. Novoselov, and A.K. Geim, Nature Phys. {\bf 2}, 620 (2006).
\bibitem{kt} T. Anto, T. Nakanishi, and R.Saito, J. Phys. Soc. Jpn. {\bf 67}, 2857 (1998); V.V. Cheianov and V.I. Fal'ko, Phys. Rev. B {\bf 74}, R041403 (2006);
C.W.J. Beenakker, Rev. Mod. Phys. {\bf 80}, 1337 (2008).
\bibitem{silvestrov07} P.G. Silvestrov, and K.B. Efetov, Phys. Rev. Lett. {\bf 98}, 016802 (2007).
\bibitem{matulis08} A. Matulis, and F.M. Peeters, Phys. Rev. B {\bf 77}, 115423 (2008).
\bibitem{apalkov08} P. Hewageegana and V. Apalkov, Phys. Rev. B {\bf 77}, 245426 (2008).
\bibitem{badarson09} J.H. Badarson, M. Titov, and P.W. Brouwer, Phys. Rev. Lett. {\bf 102}, 226803 (2009).
\bibitem{pal11} G. Pal, W. Apel, and L. Schweitzer, Phys. Rev. B {\bf 84}, 075446 (2011).
\bibitem{feshke13} R.L. Heinisch, F.X. Bronold, and H. Fehske, Phys. Rev. B {\bf 87}, 155409 (2013).
\bibitem{feshke13epl} A. Pieper, R.L. Heinisch, and H. Fehske, EPL {\bf 104}, 47010 (2013).
\bibitem{pereira06} J. M. Pereira, V. Mlinar, F.M. Peeters, and P. Vasilopoulos, Phys. Rev. B {\bf 74}, 045424 (2006).
\bibitem{fp} M.R. Masir, P. Vasilopoulos and F.M. Peeters, Phys. Rev. B {\bf 82}, 115417 (2010).
\bibitem{nanoribbons} L. Brey, and H.A. Fertig, Phys. Rev. B {\bf 73}, 235411 (2006); D.A. Areshkin, D. Gunlycke, and C.T. White, Nano Lett. {\bf 7}, 204 (2007).
\bibitem{tworzydlo} J. Tworzydlo, B. Trauzettel, M. Titov, A. Rycerz, and C. W. J. Beenakker, Phys. Rev. Lett. {\bf 96}, 246802 (2006).
\bibitem{sgmr} H. Sellier, B. Hackens, M.G. Pala, F. Martins, S. Baltazar, X. Wallart, L. Desplanque, V. Bayot, and S. Huant, Sem. Sci. Tech. \textbf{26}, 064008 (2011);  D.K. Ferry, A.M. Burke, R. Akis, R. Brunner, T.E. Day, R. Meisels, F. Kuchar, J.P. Bird, and B.R. Bennett, Sem. Sci. Tech. \textbf{26}, 043001 (2011).
\bibitem{b1} J. Berezovsky, M.F. Borunda, E.J. Heller, and R.M. Westervelt, Nanotechnology {\bf 21}, 274013 (2010).
\bibitem{b2}J. Berezovsky and R.M. Westervelt, Nanotechnology {\bf 21}, 274014 (2010).
\bibitem{b3} R. Jalilian, L. A. Jauregui, G. Lopez, J. Tian, C. Roecker, M.M. Yazdanpanah, R.W. Cohn, I. Jovanovic, and Y.P. Chen, Nanotechnology {\bf 22}, 295705 (2011).
\bibitem{c1} M.R. Connoly, K.L. Chiu, A. Lombardo, A. Fasoli, A.C. Ferrari, D. Anderson, G.A.C. Jones, and C.G. Smith, Phys. Rev. B {\bf 83}, 115441 (2011).
\bibitem{c2} A.G.F. Garcia, M. K\"onig, D. Goldhaber-Gordon, K. Todd, Phys. Rev. B {\bf 87}, 085446 (2013).
\bibitem{c3} S. Schez, J. G\"uttinger, M. Huefner, C. Stamper, K. Ensslin, and T. Ihn, Phys. Rev. B {\bf 82}, 165445 (2010).
\bibitem{pereira} J. Milton Pereira, V. Mlinar, F.M. Peeters, and P. Vasilopoulos, Phys. Rev. B {\bf 74}, 045424 (2006).
\bibitem{martino} A. De Martino, L. Dell'Anna, and R. Egger, Phys. Rev. Lett. {\bf 98}, 066802 (2007).
\bibitem{chakra} Hong-Yi Chen, Vadim Apalkov, and Tapash Chakraborty, Phys. Rev. Lett.  {\bf 98}, 186803 (2007).
\bibitem{mc} E. McCann and M. Koshino, Rep. Prog. Phys. {\bf 76}, 056503 (2013).
\bibitem{szafran} B. Szafran, Phys. Rev. B \textbf{84}, 075336 (2011).
\bibitem{waka} K. Wakabayashi, Phys. Rev. B \textbf{64}, 125428 (2001).
\bibitem{uwaga} The slight asymmetry of the plot of Fig. \ref{fwn}(c) results from the fact that the dot is not placed ideally centrally inside the cavity.
\end{thebibliography}
\end{document}